\documentclass[aps,prb,reprint,groupeaddress,showpacs]{revtex4-1}

\usepackage[utf8]{inputenc}
\usepackage[colorlinks=true,linkcolor=blue,citecolor=blue,urlcolor=blue]{hyperref}
\usepackage{amssymb}
\usepackage{amsmath}
\usepackage{amsfonts}
\usepackage{braket}
\usepackage{esdiff}
\usepackage{nicefrac}
\usepackage{graphicx}
\usepackage{xcolor}

\begin{document}


\title{Time-resolved statistics of nonclassical light in Josephson photonics}


\author{Simon Dambach}
\affiliation{Institute for Complex Quantum Systems and IQST, University of Ulm, 89069 Ulm, Germany}
\author{Björn Kubala}
\affiliation{Institute for Complex Quantum Systems and IQST, University of Ulm, 89069 Ulm, Germany}
\author{Vera Gramich}
\altaffiliation[Current address: ]{Fraunhofer-Institut f\"ur Angewandte Festk\"orperphysik,
Tullastra{\ss}e 72, 79108 Freiburg, Germany}
\affiliation{Institute for Complex Quantum Systems and IQST, University of Ulm, 89069 Ulm, Germany}
\author{Joachim Ankerhold}
\affiliation{Institute for Complex Quantum Systems and IQST, University of Ulm, 89069 Ulm, Germany}


\date{\today}

\begin{abstract}
The interplay of the tunneling transfer of charges and the emission and absorption of light can be investigated in a setup, where a voltage-biased Josephson junction is connected in series with a microwave cavity. We focus here on the emission processes of photons and analyze the underlying time-dependent statistics using the second-order correlation function $g^{(2)}(\tau)$ and the waiting-time distribution $w(\tau)$.
Both observables highlight the crossover from a coherent light source to a single-photon source.
Due to the nonlinearity of the Josephson junction, tunneling Cooper pairs can create a great variety of non-classical states of light even at weak driving.
Analytical results for the weak driving as well as the classical regime are complemented by a numerical treatment for the full nonlinear case. We also address the question of possible relations between $g^{(2)}(\tau)$ and $w(\tau)$ as well as the specific information which is provided by each of them.
\end{abstract}


\maketitle


\section{Introduction}

New possibilities offered by sources of nonconventional light are of interest for a wide range of applications. These range from secure quantum communication or teleportation to measurement, starting as early as the laser with more recent proposals exploiting squeezing properties for enhanced sensitivity and to tackle noise limitations\cite{Caves1981}.
On the other hand, superconducting circuit quantum electrodynamics devices have demonstrated the ability to create on demand any arbitrary state of light in a microwave stripline cavity, as shown by mapping out the Wigner densities of Fock states and their superposition, and of cat states in a multi-shot creation and read-out cycle\cite{Hofheinz2009}.
Here we will discuss a setup combining these ideas in exploiting a superconducting device as a continuous source of microwave light with interesting quantum-statistical properties.

Over the last years, a variety of different experimental realizations\cite{Astafiev2007,Chen2011,Pashkin2011,Hofheinz2011,Chen2014,Dykman2012} and theoretical investigations\cite{Dykman2012,Marthaler2011,Padurariu2012,Leppaekangas2013,Armour2013,Gramich2013,Kubala2014,Trif2015,Armour2015,Leppaekangas2015,Portier2015} have been dedicated to the continuous emission of photons into a cavity mode by the Cooper-pair (CP) current across  a dc-biased Josephson junction (JJ).
A microwave resonator connected in series to the junction hereby works as a well-defined electromagnetic environment absorbing the energy provided by the bias to each tunneling CP by creating photons.
The nonlinearity of the JJ translates into a complex photon creation process stretching from suppression or enhancement of higher cavity excitations to multi-photon resonances.
The excited photons finally leak from the resonator so that their mean emission rate and spectrum can be measured.\cite{Hofheinz2011}
Employing high-$Q$ superconducting resonators, far-from-equilibrium states\cite{Astafiev2007,Chen2014}, with high-photon occupations can be achieved; this can be used as a basis for analyzing the quantum-classical crossover and to investigate nonlinear effects,\cite{Armour2013,Gramich2013,Kubala2014}, such as number squeezing.
Highly attractive in this class of systems is the feature that measuring the emitted microwave radiation can indirectly yield information about the CP current and its fluctuations and vice versa.\cite{Hofheinz2011}
Generally speaking, these hybrid circuits combine observational tools, methods, and phenomena  known from the  fields of quantum optics and charge transfer physics (\emph{Josephson photonics}). Ultimate goals are to design sources for the on-demand production of single photons and highly non-classical multi-photon states (entangled photons) in the microwave up to the low-frequency terahertz regime.

In this paper, we concentrate on the bright (photonic) side of this setup and study the time-resolved statistics of photonic emission events by means of the second-order correlation function $g^{(2)}(\tau)$ and the waiting-time distribution $w(\tau)$.
Both functions are well-known tools from quantum optics\cite{Mandel1962,Carmichael1989} to study correlations of photonic emission events but have recently also been addressed to electron statistics in transport systems\cite{Brandes2008,Albert2011,Albert2012,Emary2012,Haack2014,Sothmann2014}, where relations to  other common statistical properties of electronic transport, e.g., full counting statistics or the noise spectrum, have been discussed\cite{Emary2012,Brandes2008}.

To model the circuit of a voltage-biased JJ in series with a resonator, we employ effective Hamiltonians derived  in Refs.~\onlinecite{Armour2013,Gramich2013} to describe the system close to the one- and two-photon resonances.
In conjunction with a master equation of Lindblad form modeling the impact of the environment on the system, most importantly the photon leakage, we can then study photonic correlations via the two time-dependent observables $g^{(2)}(\tau)$ and $w(\tau)$.
While the full nonlinear quantum case requires generally numerical calculations, analytical expressions can be obtained in the weak-driving regime as well as in the classical limiting case on the basis of a perturbation treatment.

Crucially, the inherent nonlinearity of the JJ enters the effective Hamiltonians as a nonlinear driving term, not as a nonlinear potential.
The impact of nonlinearities is thus not only determined by the strength of driving given by the Josephson energy $E_J$, but more importantly by tunneling matrix elements of the drive Hamiltonian between resonator states.
These depend on a dimensionless parameter $\kappa$ characterizing the importance of charge quantization of the CP current by the ratio of the total charging energy to the energy of a resonator photon. In consequence, the JJ-resonator system can reduce to a number of simpler systems, from a driven harmonic oscillator or a parametric amplifier, when the CP current becomes classical ($\kappa\rightarrow0$) to a two- or few-level system for special values of $\kappa$.

We will investigate, how the crossover between these special cases is reflected in the statistics of emitted light, changing from a coherent light source to a single-photon source, where $g^{(2)}(\tau=0)=0$. Additionally, we examine possible relations between $g^{(2)}(\tau)$ and $w(\tau)$ and use the specific physical information which is provided by each of them to reveal to what extent the system features a renewal property. This is found to be the case for the harmonic-oscillator limit ($\kappa\rightarrow0$ or weak driving) as well as the two-level system.

The paper is organized as follows. In Sec.~\ref{Sec_model}, we present our theoretical model and introduce the observables for studying correlations, the second-order correlation function $g^{(2)}(\tau)$ and waiting-time distribution $w(\tau)$.
Section~\ref{Sec_results} discusses various aspects of the time-dependent statistics of photon emission events by means of these two observables close to the one-photon resonance (Sec.~\ref{SubSec_one-photon_resonance}) and later in  (Sec.~\ref{SubSec_two-photon_resonance}) for the two-photon resonance.
Relations between $g^{(2)}(\tau)$ and $w(\tau)$ and renewal properties of the system are considered in Sec.~\ref{SubSec_relations}. Finally, Sec.~\ref{Sec_conclusions} contains the conclusions and an outlook addressing the open questions for further research.

\section{Model\label{Sec_model}}
In the following, we briefly discuss our modeling of the JJ-resonator circuit, i.e., we introduce the model Hamiltonian and a quantum master equation in Lindblad form describing the impact of the environment. On the basis of the master equation, we can calculate arbitrary time-independent correlation functions, for instance, the second-order correlation function, $g^{(2)}(\tau)$, and the waiting-time distribution, $w(\tau)$,  introduced in the last part of this section.

\subsection{Hamiltonian}
The Hamiltonian of a voltage-biased JJ-resonator circuit
\begin{eqnarray}
H=\frac{q^2}{2C}+\left(\frac{\hbar}{2e}\right)^{2} \frac{1}{2L}\phi^{2}-E_{\mathrm{J}}\cos{\left(\eta\right)}-2eV_{\mathrm{eff}}N
\label{Eq_Hamiltonian}
\end{eqnarray}
is constituted from its two subunits: the resonator described as harmonic oscillator with mass $m=(\hbar/2e)^{2}C$ and frequency $\omega_{0}=1/\sqrt{LC}$ and a JJ with Josephson energy $E_{\mathop{J}}$ and phase $\eta$ across the junction.\cite{Gramich2013} These two parts are coupled dynamically by means of an effective voltage $V_{\mathrm{eff}}=V-V_{\mathrm{res}}$, where $V$ represents the external voltage and $V_{\mathrm{res}}$ the voltage drop at the resonator. Here, the number operator $N$ counts the number of CPs which have transferred across the JJ. Both charge operator $q$ and resonator phase $\phi$ as well as number operator $N$ and junction phase $\eta$ represent sets of conjugate variables, i.e., $\left[q,\phi\right]=-2{\rm i}e$ and $\left[N,\eta\right]=-{\rm i}$, respectively.

Applying a gauge transformation on this Hamiltonian \eqref{Eq_Hamiltonian}, mapping it to a frame rotating with the \mbox{(ac-)Josephson frequency}, $\omega_{\text{J}}=2eV/\hbar$, and performing a rotating-wave approximation
\footnote{The rotating-wave approximation corresponds to the perfect cavity limit ($Q\rightarrow\infty$). The relevance of non-RWA terms is discussed in Ref.~\onlinecite{Kubala2014}. Note also (see discussion in Ref.~\onlinecite{Armour2015}) that the phase operators $\exp{(\pm{\rm i}\eta)}$ can be reduced to simple phase factors under experimentally relevant conditions.} finally yields\cite{Armour2013,Gramich2013}
\begin{equation}
H^{(p)}=\hbar\Delta n-\frac{(-{\rm i})^{p}E^{*}_{\mathrm{J}}}{2}\;:\left[\left(a^{\dagger}\right)^{p}+\left(-1\right)^{p}a^{p}\right]\frac{\mathop{J}_{p}\left(\sqrt{4\kappa n}\right)}{n^{p/2}}:
\label{Eq_Hamiltonian_RWA}
\end{equation}
close to the $p$-photon resonance, $\Delta= \omega_{0}-\omega_{\mathrm{J}}/p \approx 0$.
Normal ordering is indicated by colons. The Bessel function $\mathop{J}_{p}$ of the first kind is a function of the photonic number operator $n=a^{\dagger}a$, where $a$ and $a^{\dagger}$ represent the conventional bosonic ladder operators of the resonator. The dimensionless parameter $\kappa=E_{\mathrm{C}}/\hbar\omega_{0}=\hbar/2m\omega_{0}$ is a scale related to the granularity of charges in the circuit via its total charging energy $E_{\mathrm{C}}=2e^{2}/C$, while $E^{*}_{\mathrm{J}}=E_{\text{J}}\exp{\left(-\kappa/2\right)}$ is a renormalized Josephson energy\cite{Grabert1998}. Note that experimentally $E_{\mathrm{J}}$ is easily tunable using a superconducting quantum interference device (SQUID) configuration for the JJ, from its maximum value down to the low percentage range. The parameter $\kappa$ is basically fixed by design and tunable \emph{in situ} only to a limited extent, e.g., by accessing different resonator modes. 
First experiments found $\kappa\sim 0.1$, while new setups based on highly-inductive meta-materials consisting of SQUID arrays \cite{Jung2014} may already reach $\kappa\sim {\cal{O}}(1)$.

In case that we consider the regimes of weak Josephson couplings going along with a low photon occupation in the resonator or $\kappa\ll1$ such that $\kappa\langle n\rangle_{\mathrm{st}}\ll1$, the Bessel function can be linearized using $\mathop{J}_{p}(x)\simeq x^{p}/(p!\thinspace2^{p})$. In lowest order, i.e.,\ $J_p(\sqrt{4\kappa n})/n^{p/2}\approx \kappa^{p/2}/p!$, this then leads
to $H_0^{(p)}=\hbar\Delta n-[(-{\rm i})^{p} \kappa^{p/2}E_{\mathrm{J}}^*/(p!\thinspace2)] [(a^\dagger)^p+(-1)^p a^p]$, which realizes special cases of systems which are of particular interest in the following discussion.

For $p=1$, we obtain the Hamiltonian of a \emph{driven harmonic oscillator} 
\begin{equation}
H_{0}^{(1)}=\hbar\Delta n+\frac{{\rm i}\sqrt{\kappa}E^{*}_{\mathrm{J}}}{2}\left(a^{\dagger}-a\right)\;,
\end{equation}
with the driving term inducing coherent states. For $p=2$, one has a \emph{parametric amplifier}\cite{Walls1994} 
\begin{equation}
H_{0}^{(2)}=\hbar\Delta n+\frac{\kappa E^{*}_{\mathrm{J}}}{4}[\left(a^{\dagger}\right)^{2}+a^{2}]\; ,
\label{Eq_Hamiltonian_amplifier}
\end{equation}
where the exponentiated driving term takes the form of a squeezing operator. For $p>2$, the driving source generates multi-photon processes in the resonator and thus nonlinear optical phenomena.

As we will discuss in detail in the following, the full Hamiltonian $H^{(p)}$ also allows to engineer $N$-level systems by restricting the possible number of photon excitations in the resonator through a proper choice of the $\kappa$ parameter. For example, for $p=1$ fixing $\kappa=2$, leads to a vanishing transition matrix element $T_{m,m+1}= \Braket{m|H^{(1)}|m+1}$ for $m=1$.\footnote{Rewriting the normal-ordered Bessel function, the transition matrix elements can be expressed as \unexpanded{$T_{m,m+1}\propto\Braket{m|\exp{[{\rm i}\sqrt{\kappa}(a^{\dagger}+a)]}|m+1}$} reflecting the connection to Franck-Condon physics.} That means, the driving JJ cannot cause transitions between the neighboring excited states $1$ and $2$ of the resonator and we obtain an effective \emph{two-level system} since all inaccessible higher levels can be ignored (at zero temperature).

\subsection{Quantum master equation}

The dynamics of the reduced density operator of the system at zero temperature is described by a master equation in Lindblad form\cite{Walls1994}
\begin{equation}
\diff{\rho}{t}=\mathfrak{L}\rho=-\frac{{\rm i}}{\hbar}\left[H^{(p)},~\rho\right]+\frac{\gamma}{2}\left(2a\rho a^{\dagger}-n\rho-\rho n\right),
\label{Eq_master_equation}
\end{equation}
which can be written in terms of the Liouvillian $\mathfrak{L}$. According to the experimental realization,\cite{Hofheinz2011} the dissipator captures the effect of high-frequency modes of the electromagnetic environment acting as a dissipative heat bath on the circuit which leads to photon leakage from the resonator.
The corresponding rate $\gamma$, the inverse of the photon lifetime in the resonator, is often expressed as a quality factor $Q=\omega_{0}/\gamma$.
Experimental observations\cite{Hofheinz2011} and theoretical considerations\cite{Gramich2013} show that
additional low-frequency voltage noise affecting the JJ is weak and can be neglected for all observables considered here.

In the following, we employ a decomposition, $\mathfrak{L}=\mathfrak{L}_{0}+\mathfrak{J}$, of the time evolution of the system\cite{Brandes2008,Molmer1993,Plenio1998} into two parts; one, describing the emission of a photon from the resonator in terms of a jump operator $\mathfrak{J}$ defined by its action on the reduced density operator:
\begin{equation}
\mathfrak{J}\rho=\gamma a\rho a^{\dagger}.
\label{Eq_jump_operator}
\end{equation}
The remaining part $\mathfrak{L}_{0}$ captures the dissipative, theoretically, but deterministic system dynamics while no photon emission events occur.

\subsection{Observables for studying correlations}
Two observables will be used in the following to investigate the time-independent statistics of photon emission events. Their formal definition within the master-equation formalism is based on the Liouvillian decomposition $\mathfrak{L}=\mathfrak{L}_{0}+\mathfrak{J}$.

\emph{The second-order correlation function}
\begin{equation}
g^{(2)}(\tau)=\frac{\langle\mathfrak{J}{\rm e}^{\mathfrak{L}\tau}\mathfrak{J}\rangle_{\mathrm{st}}}{\langle\mathfrak{J}\rangle^{2}_{\mathrm{st}}}
\label{Eq_second-order_correlation}
\end{equation}
is a measure of correlations between two system jumps and thus two photon emission events separated by a time $\tau$.\cite{Mandel1962,Emary2012} The notation $\langle\dots\rangle_{\mathrm{st}}$ indicates here that the system is in steady state before the first jump, i.e., $\langle O\rangle_{\mathrm{st}}=\mathrm{Tr}\lbrace O\rho_{\mathrm{st}}\rbrace$ with $\mathfrak{L}\rho_{\mathrm{st}}=0$. For the time-independent problem here this also implies dependence on the time gap $\tau$ only (and not individually on the two jump times).
The time-evolution in-between the jumps is governed by the full Liouvillian $\mathfrak{L}$ allowing for additional jumps.

\emph{The waiting-time distribution}
\begin{equation}
w(\tau)=\frac{\langle\mathfrak{J}{\rm e}^{\mathfrak{L}_{0}\tau}\mathfrak{J}\rangle_{\mathrm{st}}}{\langle\mathfrak{J}\rangle_{\text{st}}}
\label{Eq_waiting-time_distribution}
\end{equation}
is the probability distribution for a delay $\tau$ between two subsequent system jumps and thus two subsequent photon emission events.\cite{Carmichael1989,Brandes2008} Now, the time-evolution is determined by $\mathfrak{L}_{0}=\mathfrak{L}-\mathfrak{J}$ excluding any photon emission events during the time delay $\tau$.

The waiting-time distribution according to Eqs.~\eqref{Eq_waiting-time_distribution} and \eqref{Eq_jump_operator}
is sometimes referred to as a reduced form of a `full' waiting-time distribution. 
The jump operator $\mathfrak{J}$ is based upon ladder operators $a^{(\dagger)}$, which in the current context should be understood as a sum of many `elementary transitions',  $\sum_{n}\sqrt{n+1}\ket{n}\bra{n+1}$, with each transition connecting only two states in Fock space. 
$\mathfrak{J}$ is thus a sum over several elementary jump processes, defined as resulting in a certain fixed density matrix state, and so can not resolve the nature of the transition completely.\cite{Brandes2008}
This reflects, that from the fact that a photon leaking from the cavity is detected, no conclusion can be drawn as to which particular transition took place.

Experimental observation of both $g^{(2)}(\tau)$ and $w(\tau)$ is, of course, well established in quantum optics.
In the microwave regime, correlation function measurements have been performed by the use of linear detectors\cite{Bozyigit2011,daSilva2010,Lang2011} and some progress has already been made for the setup considered here\cite{Portier2015}.

\section{Results\label{Sec_results}}

\subsection{$g^{(2)}(\tau)$ and $w(\tau)$ at the one-photon resonance\label{SubSec_one-photon_resonance}}

This first part of the results section focuses on the physics at the fundamental resonance, where the voltage applied to the JJ is tuned such that the energy of a CP tunneling across the junction equals the energy of one photon at the resonator and hence each tunneling CP excites one photon.

\subsubsection{From the harmonic oscillator to the two-level system}

\begin{figure*}
\hspace*{-0.5cm}
\includegraphics[width=1.0\textwidth]{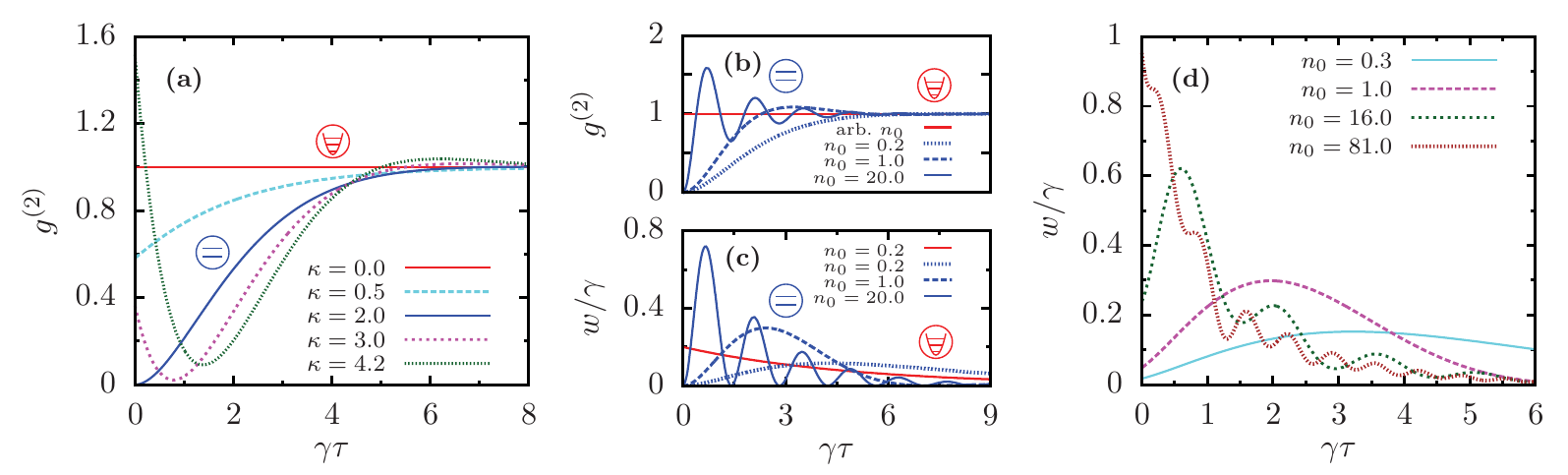}
\caption{\label{Fig_1}
(a) The second-order correlation function, $g^{(2)}(\tau)$, for weak driving, $n_{0}=0.2$. The parameter $\kappa$ allows to tune the system from a harmonic oscillator ($\kappa=0$) to a two-level system ($\kappa=2$). (b) The second-order correlation function, $g^{(2)}(\tau)$, and (c) the normalized waiting-time distribution, $w(\tau)$, for the special cases of a harmonic oscillator (red lines) and a two-level system (blue lines).
The two-level system is characterized by damped Rabi oscillations, the harmonic oscillator by a constant value [for $g^{(2)}(\tau)$], and an exponential decay with the decay rate $\gamma n_{0}$ [for $w(\tau)$]. The long-time behavior of $w(\tau)$ for the two-level system in the weak-driving regime is given by that of the harmonic oscillator.
(d) The normalized waiting-time distribution, $w(\tau)$, at $\kappa=1.5$ for various driving strengths, $n_{0}$.
In the strong-driving limit, it shows approximately the behavior of an exponential decay similar to the harmonic case, however, superimposed by oscillations.
}
\end{figure*}

Following, we will present results for the $g^{(2)}(\tau)$ and the $w(\tau)$ functions from the numerical solution of the Lindblad master equation \eqref{Eq_master_equation} and discuss various analytical approximations in certain regimes. Before, however, we will shortly recapitulate, what to expect for the special cases, where the system Hamiltonian reduces to a driven harmonic oscillator and to a driven two-level system.

\emph{For the harmonic oscillator} it is well known and easily shown using the Lindblad master equation, that in the (detuned) driven damped case its stationary solution is a coherent state $\ket{\alpha}=\exp{\left(-|\alpha|^{2}/2\right)}\sum_{q=0}^{\infty}\alpha^{q}\ket{q}/\sqrt{q!}$ with amplitude $\alpha$ fulfilling classical equations of motion
\begin{equation}
0\overset{\mathrm{st}}{=}\dot{\alpha}=-\left(\frac{\gamma}{2}+{\rm i}\Delta\right)\alpha+\frac{\sqrt{\kappa}E^{*}_{\mathrm{J}}}{2\hbar}.
\label{Eq_EOM_HO}
\end{equation}
The corresponding density matrix reflects the defining property of the coherent state and is an eigenstate of the jump operator $\mathfrak{J}$ \eqref{Eq_jump_operator} for any $\alpha$. For $\alpha$ fulfilling the steady-state equation of motion \eqref{Eq_EOM_HO} it is also an eigenstate of $\mathfrak{L}$ to eigenvalue $0$ and thus also an eigenstate of $\mathfrak{L}_0$.
The time-propagation from $t$ to $t+\tau$ under $\mathfrak{L}$ or $\mathfrak{L}_{0}$ then follows directly and yields
\begin{eqnarray}
g^{(2)}_\mathrm{HO}(\tau)&=&1,\label{Eq_g2_HO}\\
w_\mathrm{HO}(\tau)&=&\gamma\langle n\rangle_{\mathrm{st}}{\rm e}^{-\gamma\langle n\rangle_{\mathrm{st}}\tau}\label{Eq_w_HO}.
\end{eqnarray}
Both functions clearly display the Poissonian nature of the photon emission events being statistically independent.
In contrast to $g^{(2)}(\tau)$, the waiting-time distribution $w(\tau)$ depends on a mean decay rate $\gamma\langle n\rangle_{\mathrm{st}}=(\sqrt{\kappa}E^{*}_{\mathrm{J}}/\hbar\gamma)^{2}\gamma/(1+4\Delta^{2}/\gamma^{2})$ and thus is a function of the driving strength $\sqrt{\kappa}E^{*}_{\mathrm{J}}/\hbar\gamma$ as well as the detuning $\Delta$. In the following, we will generally use the stationary mean photon occupation
\begin{equation}\label{n_0}
n_{0}=\kappa \left( \frac{E^*_{\mathrm{J}}} {\hbar\gamma}\right)^{2}
\end{equation}
of the harmonic oscillator in case of resonance as a reference measure of the driving strength even in the non-harmonic regimes.

\emph{For a two-level system}, the $g^{(2)}(\tau)$ and $w(\tau)$ functions can also be straightforwardly calculated, as the time propagation under $\mathfrak{L}$ or $\mathfrak{L}_{0}$ can be easily explicitly solved for the two-level Hilbert space.
This leads to damped Rabi oscillations, e.g., on resonance  $\Delta=0$,\cite{Carmichael1989,Breuer2002}
\begin{equation}
\begin{aligned}
\!\!\! g^{(2)}_\mathrm{TLS}(\tau)&=\!1\!\!-\!{\rm e}^{-\frac{3}{4}\gamma\tau}\!\!\left[\cosh{\left(\mu_{1}\gamma\tau\right)}\!\!+\!\!\frac{3}{4\mu_{1}}\sinh{\left(\mu_{1}\gamma\tau\right)}\right] \\
\!\!\!  w_\mathrm{TLS}(\tau)&=\frac{\gamma}{\mu_{2}^{2}}n_{0}{\rm e}^{-\gamma\tau/2}\sinh^{2}{\left(\frac{\mu_{2}}{2}\gamma\tau\right)}
\end{aligned}
\end{equation}
with $\mu_{1}=\sqrt{(1/4)^{2}-n_{0}}$ and $\mu_{2}=\sqrt{1/4-n_{0}}$.

These results for the special cases of a harmonic oscillator and a two-level system, shown in the middle panel of Fig. \ref{Fig_1}, will in the following serve as a starting point to understand deviations from and the crossover between those special cases. The crossover is essentially governed by the two parameters $\kappa$ and $n_{0}$. While the former effectively changes the level structure by determining the relative size (and phase) of the driving matrix elements, the latter corresponds to the driving strength and decides how much of the nonlinearity of the system can actually be explored.

Both the $g^{(2)}(\tau)$ and the $w(\tau)$ correlation function can be expressed as an expectation value of $\mathfrak{J}$ with respect to a modified density operator $\tilde{\rho}(\tau)=\Omega(\tau)\mathfrak{J}\rho_{\mathrm{st}}$ with $\Omega(\tau)$ being the respective time-evolution operator. $g^{(2)}(\tau)$ and $w(\tau)$ can thus explicitly be calculated by means of a numerical implementation of the time evolution of the reduced density operator $\rho$
governed by $\mathfrak{L}$ and $\mathfrak{L}_{0}$, respectively.

Numerical results are displayed in the left and right panels of Fig.~\ref{Fig_1}, setting $\Delta=0$ for simplicity. While Fig.~\hyperref[Fig_1]{1(a)} shows how the $g^{(2)}(\tau)$ changes with $\kappa$ for a fixed small driving strength $n_{0}=0.2$, Fig.~\hyperref[Fig_1]{1(d)} pictures the $w(\tau)$ function for fixed $\kappa=1.5$ and increasing driving strength.

We easily recognize the two special cases of the harmonic oscillator and the two-level system additionally labeled by the respective pictogram in the plot of $g^{(2)}(\tau)$. Due to the weak driving and the corresponding small $\langle n\rangle_{\mathrm{st}}$, only the lowest states are occupied and the system is close to equilibrium.
$g^{(2)}(\tau)$ approaches here its long-time limit of $1$ very soon featuring only small damped oscillations around this value.
Since the $w(\tau)$ function represents a normalized probability distribution, it will vanish in the long-time limit.
As in the harmonic-oscillator case \eqref{Eq_w_HO}, $w(\tau)$ includes exponential decay terms with rates $\sim \gamma  n_{0}$.
In the limit of strong driving, this exponential decay, however, is superimposed by oscillations with frequency $\sim\gamma\sqrt{n_{0}}$.
For weak driving, the short-time behavior of both $g^{(2)}(\tau)$ and $w(\tau)$ is governed by a time scale $\sim\gamma^{-1}$.

More features of the full numerical results will be explained by the analytical solution in the weak-driving limit (see Sec.~\ref{SubSubSec_weak-drive}), where we will also discuss Fig.~\hyperref[Fig_1]{1(a)} further. As a preliminary conclusion, we can state, that our system realizes a harmonic oscillator for $\kappa=0$ and a two-level system for $\kappa=2$. For arbitrary values of $\kappa$, correlations  for  the weakly driven system are similar to the harmonic case after some initial short-time deviations, while for stronger driving features resembling the two-level system appear. To understand the physical information contained in the two correlation functions in more detail, we now
turn to analytical results providing exact expressions for various contributing terms and timescales.

\subsubsection{Approximations for weak-driving and semiclassical limits\label{SubSubSec_weak-drive}}

For the numerical results above, $g^{(2)}(\tau)$ and $w(\tau)$ are found by calculating the time evolution over a time $\tau$ of the stationary density matrix after a jump occurred at time $t=0$ under the action of the Liouvillian
$\mathfrak{L}$ [or $\mathfrak{L}_0$ respectively for $w(\tau)$]. For analytical results, we similarly consider the action of the adjoint Liouvillian $\mathfrak{L}^{\dagger}$ or $\mathfrak{L}_{0}^{\dagger}$ on system operators.  Making use of the quantum regression theorem,\cite{Breuer2002} this yields a linear system of first-order differential equations to determine $g^{(2)}(\tau)$ or $w(\tau)$. However, in general, these systems will not close but produce a hierarchy of higher-order operator expressions.

A closed system of equations will appear either if the number of system states is limited, or if higher-order expressions drop out due to an (approximate) factorization into low-order terms.
The number of states will thereby be limited if transitions to higher states are forbidden due to vanishing matrix elements of the driving Hamiltonian (as discussed above in the two-level case) or small as in the limit of weak driving, $n_{0}\rightarrow0$. In contrast, higher-order expressions do not appear in the harmonic-oscillator case or will approximately drop out in the semiclassical regime ($n_{0}\gg1$, $\kappa\rightarrow0$).

\begin{figure}
\centering
\hspace*{-1.3cm}
\includegraphics{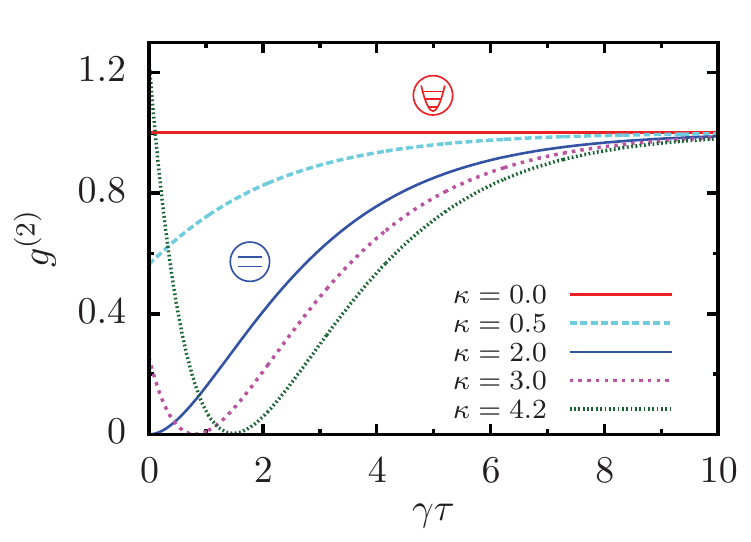}
\caption{\label{Fig_2}
The second-order correlation function, $g^{(2)}(\tau)$, in the numerical weak-driving/few-photon limit, $n_0=(\sqrt{\kappa}E^{*}_J/\hbar\gamma)^2\rightarrow 0$, where it approaches the analytical result given by Eq.~\eqref{Eq_g_2_weak_driving}, for various values of $\kappa$
(with $\kappa=0$ corresponding to the harmonic oscillator and $\kappa=2$ to a two-level system).
The nonlinearity for finite $\kappa$ becomes apparent in suppressed (enhanced) transitions to the second-excited state for $\kappa<4\; (\kappa>4)$, resulting in $g^{(2)}(0)=(1-\kappa/2)^2$. Compare to Fig.~\hyperref[Fig_1]{1(a)} for the effects of stronger driving (see main text).
}
\end{figure}

\emph{In the regime of weak driving}, $n_0\ll1$, we can approximately consider a three-level system and thus neglect all operator contributions $\left(a^{\dagger}\right)^{m}a^{n}$ with $m>2$ or $n>2$ to find a closed system of differential equations. Solving this system then yields for the second-order correlation function the analytical expression
\begin{equation}
g^{(2)}(\tau)=1+\frac{\kappa^{2}}{4}{\rm e}^{-\gamma\tau}-\kappa\cos{\left(\Delta\tau\right){\rm e}^{-\gamma\tau/2}},
\label{Eq_g_2_weak_driving}
\end{equation}
with small next-order correction terms of order $n_{0}$.

Considering for the moment the resonant case, $\Delta=0$, the correlation takes the simple form $g^{(2)}(\tau)= (1-\kappa{\rm e}^{-\gamma\tau/2}/2)^2$ so that one observes complete anti-bunching at times $\gamma\tau=2 \operatorname{ln}(\kappa/2)$ for $\kappa\geq 2$. This is indeed found in the numerical data of Fig.~\ref{Fig_2} for extremely weak driving, while this feature tends to be absent for somewhat larger driving [see Fig.~\hyperref[Fig_1]{1(a)}]. 
As has already been discussed elsewhere,\cite{Kubala2014} in the weak driving limit $g^{(2)}(\tau=0)$ is directly linked to the occupation of the second excited state $P_2$ and, hence, to the probability that the resonator gets excited by a second CP tunneling event before it has relaxed to its ground state. This occupation can be found by a simple rate-equation description resulting in
\begin{equation}\label{g2(0)}
 g^{(2)}(0)\approx \frac{2P_2}{P_1^2}\approx\frac{1}{2} \left|\frac{T_{1,2}}{T_{0,1}}\right|^2 = \left(1- \frac{\kappa}{2}\right)^{2}\;,
\end{equation}
where in this regime of Coulomb-blockade a $P(E)$-type calculation\cite{Grabert1992} links the golden-rule excitation rates to the absolute values of the transition matrix elements of the driving Hamiltonian\cite{Hofheinz2011}. Anti-bunching, $g^{(2)}(0)<1$, observed in the range of $0<\kappa<4$ is thus the effect of the suppression of higher-order excitations due to the nonlinearity of the system [as reflected in the normal-ordered Bessel function in Eq.~\eqref{Eq_Hamiltonian_RWA}] on the few photon level.

While the anti-bunching, $g^{(2)}(0)<1$, an indicator of the quantized, non-classical nature of the cavity's electromagnetic field can be derived from an incoherent $P(E)$ description, it turns out that the dynamical features in $g^{(2)}(\tau)$ can not be gained such, but only in a full quantum-mechanical picture of coherent excitations. In particular, the short-time behavior $g^{(2)}(\tau)\lessgtr g^{(2)}(0)$ for $\kappa\gtrless2$ can be traced back to the relative phase of certain transition matrix elements.

Contrasting Fig.~\ref{Fig_2} to Fig.~\hyperref[Fig_1]{1(a)} highlights the effects of somewhat stronger driving:\\ (i) the suppression (enhancement) of $g^{(2)}(0)$ for $\kappa<4$ ($\kappa>4$) is reduced (enlarged) as higher oscillator states contribute, while $g^{(2)}(0)\equiv 0$ for $\kappa =2$ for any driving strength,\\
(ii) the dip occurring in $g^{(2)}(\tau)$ after a short time (for $\kappa \gtrsim 2$) is reduced with $g^{(2)}(\tau)>0$ for all times and\\
(iii) oscillations with $g^{(2)}(\tau)> g^{(2)}(\infty)=1$ develop.

Employing the corresponding approximation scheme for the weak-driving regime, the waiting-time distribution in leading order (and again for $\Delta=0$) is calculated
\begin{equation}
\frac{w(\tau)}{\gamma n_{0}}={\rm e}^{-\gamma n_{0}\tau}+\frac{\kappa^2}{4}{\rm e}^{-\gamma\tau}-\kappa {\rm e}^{-\gamma\tau/2}
\label{Eq_w_weak_driving}
\end{equation}
with small next-order correction terms of order $n_{0}$.
Comparing to Eq.~\eqref{Eq_g_2_weak_driving}, the close relation between $g^{(2)}(\tau)$ and $w(\tau)$ for weak driving is apparent.

Apart from their different normalization, we observe that the second and third terms of both expressions with the decay rates $\gamma$ and $\gamma/2$ [and thus the short-time behavior of  $g^{(2)}(\tau)$ and $w(\tau)$] coincide.
This happens since the underlying systems of differential equations based on $\mathfrak{L}$ and $\mathfrak{L}_{0}$ differ in higher-order operator expressions which do not contribute in leading order in $n_{0}$ to these two terms (but show solely an effect on the first term).
Nonetheless, for long times $w(\tau)$ finally  shows a slow exponential decay on a time scale $1/(\gamma n_{0})$, while $g^{(2)}(\tau)\rightarrow 1$ in this limit.

Notably, the time scale $(\gamma/2)^{-1}$ which impacts the time-dependent behavior of both $g^{(2)}(\tau)$ and $w(\tau)$ in the strongly overdamped case can be traced back to the decay rate of coherences, i.e., off-diagonal entries of the density matrix. This, in turn, means, that in contrast to $g^{(2)}(\tau=0)$ \eqref{g2(0)} \emph{the time dependence of the correlation functions can not be properly described within a $P(E)$-like rate-equation approach even in the weak-driving limit}. This also highlights the profound difference of the photonic two-level case in comparison to the case of correlation functions of (spinless) electrons transported through a single site, where coherences do not matter\cite{Brandes2008,Emary2012}.

\emph{In the semiclassical regime},  the operators $a^{(\dagger)}$ can be replaced by $\alpha^{(*)}+\delta a^{(\dagger)}$, with $\delta a^{(\dagger)}$ describing quantum fluctuations around the classical solution of the complex oscillation amplitude $\alpha^{(*)}=\langle a^{(\dagger)}\rangle$. In the semiclassical regime, fluctuations will then be small compared to the mean amplitude and operator contributions beyond linear order can be neglected.

Such a semiclassical approach proved to be very fruitful in calculating stationary expectation values revealing, for instance, bifurcations between different solutions.\cite{Armour2013,Armour2015}

In principle, one can proceed similarly here and calculate $g^{(2)}(\tau)$ and $w(\tau)$ in the semiclassical limit. It turns out, however, that in the regime, where the semiclassical approximation is valid, the resulting behavior of both functions is hardly distinguishable from  that of a harmonic oscillator:  $g^{(2)}$ is basically $1$ [Eq.~\eqref{Eq_g2_HO}] with tiny corrections of order $1/|\alpha|^{2}$ and $w(\tau)$ is dominated by a strong exponential decay [Eq.~\eqref{Eq_w_HO}] due to a large $\langle n\rangle_{\mathrm{st}}$, which completely obscures small (quantum) corrections occurring in the long-time limit.

\subsubsection{Effective $N$-level systems}

\begin{figure}
\centering
\hspace*{-0.3cm}
\includegraphics{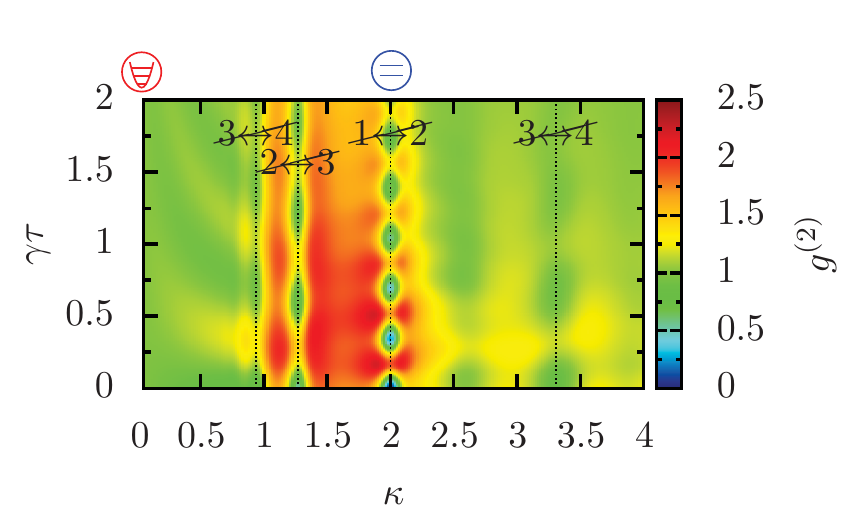}
\caption{\label{Fig_3}
The second-order correlation function, $g^{(2)}(\tau)$, for strong driving, $n_0=18^2$. At special values of $\kappa$ (dotted lines) certain transition matrix elements vanish, reducing the number of involved cavity states. This yields pronounced (damped) oscillations in $g^{(2)}(\tau)$ in a certain $\kappa$ interval around these values.
Lines below $\kappa=2$ can be traced back to the first zero at $3.83$
of the Bessel function $J_{1}$, e.g., $T_{2,3}\propto\Braket{2|: J_{1}(\sqrt{4\kappa n})/\sqrt{n}:|2}=0$ for $\kappa=1.27$, the line at $\kappa=3.31$ to the second zero.
}
\end{figure}

Above, we explained, how for the special value of $\kappa=2$ the system is reduced to a two-level system. Analyzing the roots of higher transition matrix elements between neighboring states reveals the occurrence of other few-level systems for specific choices of $\kappa$. (See Ref.~\onlinecite{Bretheau2015}, where $N$-level systems are realized by dynamically blocking transitions with an additional external drive.) For example, the conditions $T_{2,3}=0$ valid for $\kappa=3\pm\sqrt{3}$ and $T_{3,4}=0$ valid for $\kappa\approx0.94$, $3.31$, or $7.76$ lead to effective three- and four-level systems, respectively.

Note that the vanishing of transition matrix element $T_{m,m+1}\propto\Braket{ m|:J_{1}(\sqrt{4\kappa n})/\sqrt{n}:|m}$ corresponds to zeros of the Bessel function $J_1$ for large $m$. Due to normal ordering, however, for small $m$ the roots are shifted, e.g., for $m=1$ we have
$\Braket{1|~J_{1}(\sqrt{4\kappa n})/\sqrt{n}~|1}= J_{1}(\sqrt{4\kappa})=0$ for $\kappa=3.67$ without the normal ordering instead of $\kappa=2$.

Figure~\ref{Fig_3} displays the second-order correlation function as a contour plot in the regime of very strong driving ($n_{0}=18^{2}$) for $\kappa\in\left[0,~4\right]$. We immediately recognize the prominent Rabi oscillations in the case of a two-level system, see Fig.~\hyperref[Fig_1]{1(b)}. However, also $N$-level systems feature Rabi-like oscillations indicating those special values of $\kappa$ where higher transitions are suppressed. The amplitude as well as the frequency of these oscillations decrease with an increasing number of remaining levels. Notably, the Josephson-driven resonator setup thus allows by properly adjusting the $\kappa$ parameter to turn the harmonic oscillator into a highly nonlinear $N$-level system. If $\kappa$ could be varied \emph{in situ} (for example by an external magnetic flux in case of a SQUID array), one could even switch from $N$ to $N'$ level systems on short-time scales.

\subsubsection{Detuning from the one-photon resonance}

\begin{figure}
\centering
\hspace*{-0.3cm}
\includegraphics{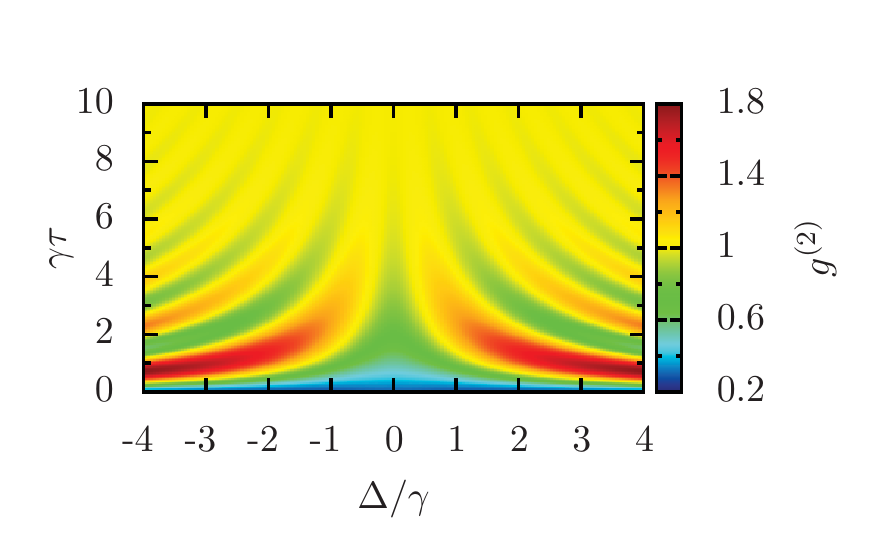}
\caption{\label{Fig_4}
The second-order correlation function, $g^{(2)}(\tau)$, as a function of detuning, $\Delta$, at $\kappa=1$ for weak driving, $n_{0}=0.2$.
Oscillations with frequency $f\approx\Delta/2\pi$ are observable with an amplitude, which increases with increasing detuning.
Compare to the analytical result given by Eq. \eqref{Eq_g_2_weak_driving} for the effects occurring in the weak-driving limit, $n_{0}=(\sqrt{\kappa}E^{*}_{\mathrm{J}}/\hbar\gamma)^{2}\rightarrow0$.
}
\end{figure}

We drop now the restriction of $\Delta=0$ to study the effect of detuning on the system. Considering the special case of a harmonic oscillator, the analytical results for both $g^{(2)}(\tau)$ [Eq.~\eqref{Eq_g2_HO}] and $w(\tau)$ [Eq.~\eqref{Eq_w_HO} are still valid without limitations but it should be noted that $\langle n\rangle_{\mathrm{st}}$ is modified by a factor of $1/[1+4(\Delta/\gamma)^{2}]$, thus leading to a much slower decay for high detuning.
Leaving this special case and going to higher $\kappa$, the $g^{(2)}$ and $w(\tau)$ functions are both superimposed by oscillations. In the regime of weak driving, the behavior of $g^{(2)}(\tau)$ is directly given by Eq.~\eqref{Eq_g_2_weak_driving} revealing oscillations with frequency $\omega=\Delta$ and high amplitudes for strong detuning, whereas $g^{(2)}(0)$ remains unchanged. The numerical results of $g^{(2)}(\tau)$ for somewhat stronger driving ($n_{0}=0.2$) in Fig.~\ref{Fig_4} show small differences to the analytical results, only.
Note that the far off-resonant two-level system also shows Rabi oscillations with frequency $\Delta$ for strong driving.

\subsection{Relations between $g^{(2)}(\tau)$ and $w(\tau)$\label{SubSec_relations}}

In order to investigate the statistics of the photon emission events in various regimes of our system, we have so far discretionarily elected to discuss at times the second-order correlation function $g^{(2)}(\tau)$ or the waiting-time distribution $w(\tau)$. Obviously, this immediately throws up the question, to what extent the information provided by $g^{(2)}(\tau)$ and $w(\tau)$ differ or complement each other.

Apparently, both tools oftentimes provide very similar information and are closely linked. Consider, for instance, that their definition [Eqs.~\eqref{Eq_second-order_correlation} and \eqref{Eq_waiting-time_distribution}] directly yields the equal-time relation $w(0)=\gamma\langle n\rangle_{\mathrm{st}}g^{(2)}(0)$, and that we have observed similar Rabi-type oscillations in both functions, when certain transitions between excited neighboring states are suppressed. 

In general, however, there is no one-to-one correspondence between $g^{(2)}(\tau)$ and $w(\tau)$. In the limit, where
each relaxation process resets the system to a known state, a situation described in this context by renewal theory, such an equivalence exists though. To see that, one works with characteristic second-order Fano factors (see following) for the respective quantities.

\begin{figure*}
\centering
\includegraphics{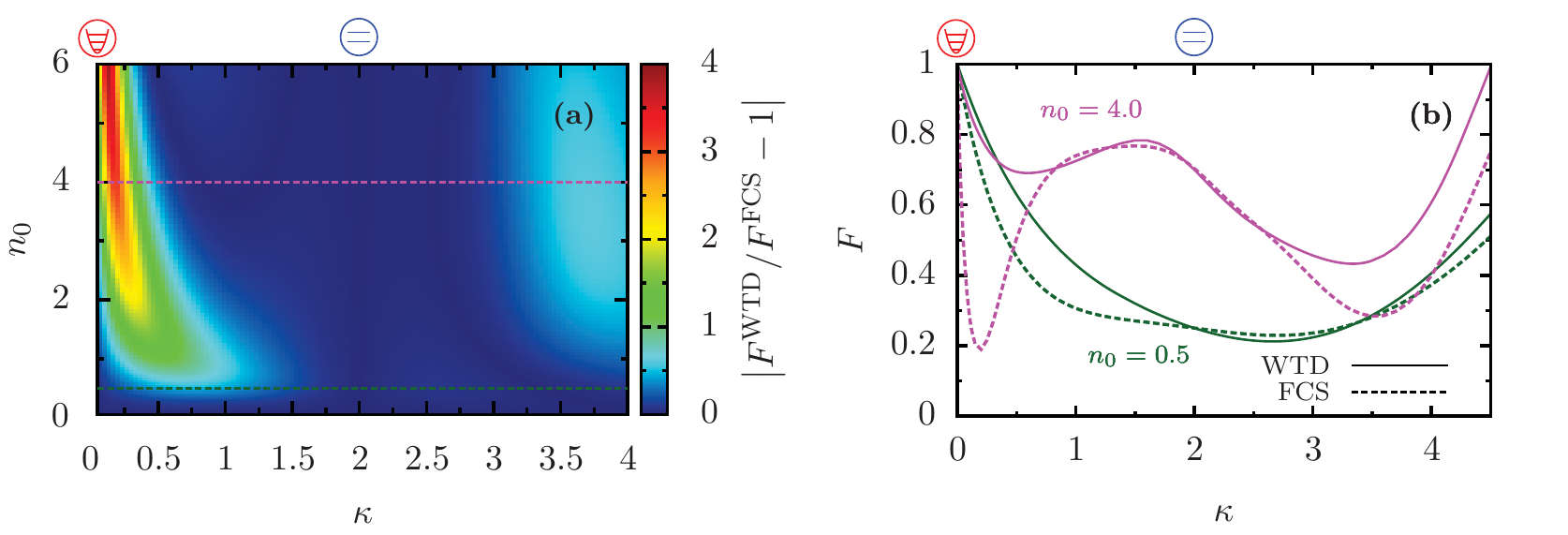}
\caption{\label{Fig_5}
In case that renewal theory applies to the system, $g^{(2)}(\tau)$ and $w(\tau)$ provide identical information and the two Fano factors deduced from them are identical: $F^{\text{FCS}}=F^{\text{WTD}}$.
The measure $|F^{\text{WTD}}/F^{\text{FCS}}-1|$ (a) highlights the deviations of the system from renewal theory, which vanish at $\kappa=0$ (harmonic oscillator), $\kappa=2$ (two-level system), and $n_{0}\ll1$ (weak-driving regime, where nonlinearities are not crucial). The dotted lines refer to two cross sections (b) at weak and strong driving, revealing sub-Poissonian distributions of emitted photons.
}
\end{figure*}

\subsubsection{Renewal theory}

A Poissonian process, where events are occurring independently, can alternatively be considered as characterized by the fact that after each event the probability for the subsequent event is given by an identical, independent exponential probability distribution. Renewal theory generalizes this to so-called renewal processes characterized by an identical, independent, but otherwise arbitrary probability distribution.\cite{Cox1995} Hence, each event resets the system to the very same state and thus renews it.

For such a renewal process, the two functions $g^{(2)}(\tau)$ and $w(\tau)$ can be directly related (in Laplace space),\cite{Emary2012} and thus provide identical information.
From that relation a somewhat simpler comparison restricted to low-order moments follows.\cite{Albert2011,Albert2012,Haack2014} Indicating the relative strength of second-order fluctuations, a single, dimensionless number, the Fano factor is used.

The standard Fano factor, $F^{\text{FCS}}$, characterizing the statistics of the photon flux $I_\mathrm{ph}$ leaking from the cavity, can in the spirit of \emph{full counting statistics} (FCS) be defined from the statistic of the number of photons leaked during an accumulation time, $N_T=\int_0^T d\tau I_\mathrm{ph}(\tau)$ as
\begin{equation}
F^{\text{FCS}}=\left. \frac{\langle N^{2}_T\rangle-\langle N_T\rangle^{2}}{\langle N_T\rangle}\right|_{T\rightarrow \infty}\,.
\label{Eq_Fano_factor_FCS_basic}
\end{equation}
Following Ref.~\onlinecite{Mandel1962} it can also be gained from $g^{(2)}(\tau)$,
\begin{eqnarray}
F^{\text{FCS}}
&=&1+2\langle I_\mathrm{ph}\rangle\int^{\infty}_{0}\,\mathrm{d}\tau\left[g^{(2)}(\tau)-1\right]\label{Eq_Fano_factor_FCS}\\
&=&\frac{S(\omega=0)}{2\langle  I_\mathrm{ph} \rangle}\nonumber \,,
\end{eqnarray}
where the second line highlights an alternative access to this quantity via the zero-frequency noise, $S(\omega=0)$, of the photon current.

An alternative Fano-type factor,  $F^{\text{WTD}}$, \emph{can be defined from the first two cumulants of the waiting-time distribution},
\begin{equation}
F^{\text{WTD}}=\frac{\langle\tau^{2}\rangle-\langle\tau\rangle^{2}}{\langle\tau\rangle^{2}}.
\label{Eq_Fano_factor_WTD}
\end{equation}
Then, \emph{for any renewal process} $F^{\text{FCS}}=F^{\text{WTD}}$ holds with $F^{\text{FCS}}=1=F^{\text{WTD}}$ for the Poissonian case.\footnote{Note that the stationary occupation of the cavity can be characterized by yet another type of Fano factor, \unexpanded{$(\langle n^{2}\rangle_{\mathrm{st}}-\langle n\rangle^{2}_{\mathrm{st}})/\langle n\rangle_{\mathrm{st}}$} (see Ref.~\onlinecite{Armour2013}).}
For our JJ-resonator system it is obvious that in the two-level case, $\kappa=2$, the system is reset to the very same state, namely $| 0\rangle\langle 0|$, by any jump process and, hence, the photon emission process constitutes a renewal process.

For the harmonic oscillator, realized for $\kappa \rightarrow 0$ or in the weak-driving limit, the jump operator acting on an arbitrary system density matrix does not actually reset the system to a single state. It does so, however, for any density matrix appearing during the time evolutions within the definitions [Eqs.~\eqref{Eq_second-order_correlation} and \eqref{Eq_waiting-time_distribution}] of  $g^{(2)}(\tau)$ and $w(\tau)$ , since the time evolutions start from the stationary state of the harmonic oscillator, an eigenstate of $\mathfrak{J}$ and of both $\mathfrak{L}$ and $\mathfrak{L}_{0}$. Thus, for our purposes the harmonic oscillator behaves as a single-reset system and undergoes a renewal process.

\subsubsection{JJ-resonator system and its renewal character}

In order to capture the system's deviations from renewal character, the two different definitions of Fano factors, $F^{\text{FCS}}$ and $F^{\text{WTD}}$, are compared in Fig.~\ref{Fig_5}.
If the two Fano factors differ, $g^{(2)}(\tau)$ and $w(\tau)$ are not directly linked and provide, in principle, different information about the system.
In case of a renewal process, the difference vanishes, which can be observed in Fig.~\hyperref[Fig_5]{5(a)} in a small region around
the special cases of $\kappa=0$ and $\kappa=2$ for arbitrary driving strength. It also holds true for arbitrary $\kappa$ in the regime of weak driving, where the dynamics of the system are hardly influenced by the nonlinearities of the Bessel function. Strong discrepancies from a renewal character appear in the classical regime at small but finite $\kappa$ and strong driving $n_{0}$.
The shape of that feature, visible in  Fig.~\hyperref[Fig_5]{5(a)}, reflects that nonlinearities appear, once the argument of the Bessel function becomes of order $1$, $\sqrt{4\kappa n}\sim 1$.

The two cross sections at $n_{0}=0.5$ and $4.0$ in Fig.~\hyperref[Fig_5]{5(b)} showcase two facts.
First, we observe, that there may be additional parameter values with $F^{\text{FCS}}=F^{\text{WTD}}$ while renewal theory is \emph{not} expected to be valid. Indeed, for these values, one can check, that the full time dependence of $g^{(2)}(\tau)$ and $w(\tau)$ does not give identical information, but differences only become apparent in higher moments. Secondly, note that the sub-Poissonian distribution of emitted photons in the long-time limit with $F^{\text{FCS}}<1$ for $\kappa \gtrsim 4$ does not correspond to anti-bunched photons, since, in fact, $g^{(2)}(0)>g^{(2)}(\tau)$ at all times for these values of $\kappa$ (and weak to moderate driving) [see Figs.~\hyperref[Fig_1]{1(a)} and \ref{Fig_2}].

Furthermore, we observe $F^{\text{FCS}}>1$ for strong driving, e.g., choosing $n_{0}=18^{2}$, at $\kappa=3-\sqrt{3}$ corresponding to an effective three-level system, although we clearly expect here an anti-bunching behavior since $g^{(2)}(0)<g^{(2)}(\tau)$ at all times (see also Fig.~\ref{Fig_3}). These findings are in line with recent results \cite{Singh1983,Zou1990,Emary2012,Kronwald2013} emphasizing that Eq.~\eqref{Eq_Fano_factor_FCS} does not in general imply a one-to-one correspondence of sub-/super-Poissonian FCS and short-time (anti-)/bunching.

\subsection{$g^{(2)}$ and $w(\tau)$ at the two-photon resonance\label{SubSec_two-photon_resonance}}

\begin{figure}
\centering
\hspace*{-1.3cm}
\includegraphics{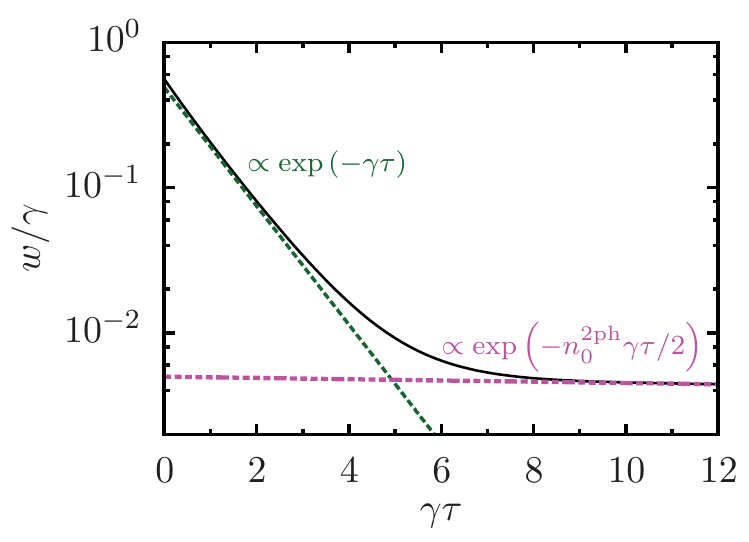}
\caption{\label{Fig_6}
The waiting-time distribution (solid line), $w(\tau)$, for the two-photon resonance in the special case of a parametric amplifier, $\kappa=0.001$, $n^{\mathrm{2ph}}_{0}=0.02\ll1$, which approximately coincides with the analytical result given by Eq.~\eqref{Eq_w_2ph}. The short- and long-time decays are indicated by the analytical results (dashed lines), corresponding to the probability of detecting the first or second photon of an emitted pair.
}
\end{figure}

So far, only small detuning around the resonance condition $\omega_{\mathrm{J}}=\omega_{0}$ for one-photon processes have been investigated. Tuning the bias voltage, other than this fundamental resonance can be accessed.
In particular, we now turn to the simplest higher-order creation processes, the two-photon process with resonance condition $2\omega_{0}\approx\omega_{\mathrm{J}}$. We start our considerations on the basis of the Hamiltonian given in Eq.~\eqref{Eq_Hamiltonian_RWA} with $p=2$, which is in a frame rotating with $\omega_{\mathrm{J}}/2$ and within a rotating-wave approximation. For most of the following, we consider the regime of weak driving and consequently small occupation $\langle n\rangle_{\mathrm{st}}$, where the Hamiltonian simplifies to that of the parametric amplifier, $H_0^{(2)}$,  in Eq.~\eqref{Eq_Hamiltonian_amplifier}.

In this weak-driving regime, incoherent CP transport across the JJ can be observed, i.e., the events of CP tunneling processes are statistically independent and follow a Poissonian probability distribution. This is based on the fact that photon relaxation at the resonator occurs sufficiently fast compared to CP tunneling. Therefore, we always observe a pair of photons followed by a long waiting time until the next emission event occurs.

By elementary considerations following this picture, the two different  Fano factors  $F^{\text{FCS}}\neq F^{\text{WTD}}$ are found to differ, since renewal theory does not hold.
The observation that for well-separated bunches of $p$ particles the Fano factor gives the bunch size, $F^{\text{FCS}}=p$, is well known (see, e.g., Refs.~\onlinecite{Jehl2000,Lefloch2003,Belzig2005,Koch2005}).
Formally, one may argue that in the long-time limit, the number of photons $N_T=\int_0^T d\tau I_\mathrm{ph}(\tau)$ is well approximated by a coarse-graining over typical time scales of the bunch by $I_\mathrm{ph}(\tau) \approx p I_\mathrm{CP}(\tau)$, where the CP (number) current $I_\mathrm{CP}$ captures the underlying (Poissonian) statistics of the bunches. The Fano factor then immediately follows from Eq.~\eqref{Eq_Fano_factor_FCS_basic}, $F^{\text{FCS}}=\frac{p^2}{p} F^{\text{FCS}}_\mathrm{CP}=p$ with corrections determined by the separation of the typical time-scales of the bunch length and the time between bunches.

The first and second moments of the waiting-time distribution, and hence $F^{\text{WTD}}$, are found averaging  over the equally probable cases, that  the first or the second photon of a bunch (for $p=2$) is observed at $t=0$.
This gives $\langle\tau_{\mathrm{ph}}\rangle=[\mathcal{O}(\gamma^{-1})+\langle\tau_{\mathrm{CP}}\rangle]/2$ and $\langle\tau^{2}_{\mathrm{ph}}\rangle= [\mathcal{O}(\gamma^{-1})+ \langle\tau^{2}_{\mathrm{CP}}\rangle]/2=4\langle\tau_{\mathrm{ph}}\rangle^{2}$ with corrections on the order of the bunch length, $\gamma^{-1}$, resulting in $F^{\text{WTD}}=3$ (for general $p$, one similarly finds $F^{\text{WTD}}=2p-1$).

Going beyond these preliminary considerations, analytical expressions for the full time-dependence  of $g^{(2)}(\tau)$ and  $w(\tau)$ can again be found in the weak-driving limit (restricted to the resonant case,  $\omega_{0}=\omega_{\mathrm{J}}/2$, again).

The instantaneous limit,
\begin{equation}
g^{(2)}(0)=\frac{1}{2n^{\mathrm{2ph}}_{0}}+2
\end{equation}
with next correction terms of the order of the stationary mean occupation number $n_{0}^{\text{2ph}}=(\kappa E^{*}_{\mathrm{J}}/\hbar\gamma)^{2}/2$ (in leading order), which again measures  the driving strength, has been extensively discussed elsewhere.\cite{Kubala2014,Gramich2013,Leppaekangas2013}
We want to emphasize, that the $1/2n^{\mathrm{2ph}}_{0}$-divergence of photonic correlations  in the weak-driving limit due to the two-photon creation process can again be understood on the basis of excitation and decay rates for double and single occupancy of the cavity, while the time-dependence of $g^{(2)}(\tau)$ can not be gained correctly in this simple manner.
Note also, that depending on the quality factor of the cavity, two-photon processes may dominate the photonic correlations even around the fundamental resonance, where they compete with the resonant single-photon processes.\cite{Kubala2014}

The waiting-time distribution, shown in Fig.~\ref{Fig_6}, takes the form
\begin{equation}
\frac{w(\tau)}{\gamma}=\frac{1}{2}{\rm e}^{-\gamma\tau}+\frac{n^{\text{2ph}}_{0}}{4}{\rm e}^{-n^{\text{2ph}}_{0}\gamma\tau/2}+\frac{13n^{\mathrm{2ph}}_{0}}{4}{\rm e}^{-2\gamma\tau}
\label{Eq_w_2ph}
\end{equation}
with both coefficients and decay rates in leading order in $n^{\mathrm{2ph}}_{0}$. The expectation value $\langle \tau\rangle$
for the waiting time is dominated by the first and the second contributions, while the third one does hardly contribute for weak driving when $n_{0}^{\text{2ph}}\ll 1$. Physically, the first contribution, describing the case that the first photon of a bunch is observed at $\tau=0$, determines the waiting-time for the second photon within the same bunch. Instead, the second contribution is related to the waiting-time for the next bunch to arrive, when the second photon was observed initially.

Using the analytical results of $g^{(2)}(\tau)$ and $w(\tau)$ in conjunction with Eqs.~\eqref{Eq_Fano_factor_FCS} and \eqref{Eq_Fano_factor_WTD} results in $F^{\text{FCS}}=2+7n^{\mathrm{2ph}}_{0}/2$ and $F^{\text{WTD}}=3+3n^{\mathrm{2ph}}_{0}$ confirming our above considerations.

\section{Conclusions and outlook\label{Sec_conclusions}}

We have analyzed the time-resolved statistics of photon emission events of an electromagnetic oscillator coupled to a voltage-biased Josephson junction by means of the second-order correlation function $g^{(2)}(\tau)$ and the waiting-time distribution $w(\tau)$.
Starting from a time-independent Hamiltonian in rotating-wave approximation and the quantum master equation in Lindblad form, we have obtained numerical results as well as analytical expressions on the basis of perturbative approaches in the weak-driving and the semiclassical regime. The photonic statistics are basically governed by the parameters $E_{\mathrm{J}}/\hbar\gamma$ measuring the driving strength and $\kappa$ describing the impact of charge quantization and determining magnitude and phase of the transition matrix elements of the drive.

Changing parameters allows the realization of various simple systems and thus the observation of the photon statistics of a driven harmonic oscillator, a two-level system or a parametric amplifier, all within the voltage-biased JJ-resonator compound. Increasing the impedance and hence $\kappa$, either by design or \emph{in situ} by use of highly inductive metamaterials, the crossover from the harmonic oscillator ($\kappa\rightarrow0$) to the two-level system ($\kappa=2$) can be studied. For other special values of $\kappa$, further $N$-level systems can be implemented, exhibiting for instance Rabi-type oscillations in $g^{(2)}(\tau)$ and $w(\tau)$ for strong driving. 

While the scenarios above occur at the fundamental resonance where each tunneling Cooper pair creates one photon, higher resonances can be accessed when the voltage bias provides the energy necessary to create several photons. Then, for example, at the two-photon resonance, parametric-amplifier physics can be investigated in the $\kappa\rightarrow0$ limit. 

Photon statistics will show highly characteristic signatures in these different scenarios: most dramatically, single-photon creation and complete anti-bunching are observed for the two-level system, photon-pair creation, and bunching for the parametric amplifier. To further study the full time-resolved photon statistics, we explored the second-order correlation function $g^{(2)}(\tau)$, the waiting-time distribution $w(\tau)$, and Fano-type factors extracted from them. 
The rich variety of strongly correlated and non-classical states of light, observable even for weak driving, all stem in essence from the inherent nonlinearity of the Josephson junction appearing as a normal-ordered Bessel function in the effective RWA Hamiltonian. 

Extending our study to several modes within one or several cavities will be of particular interest for further research. If the energy of a CP transfer equals the sum of the energies of two photons, entangled photon pairs can be created. The statistics of these photon pairs have so far only been investigated by means of the second-order correlation function and Fano factor,\cite{Armour2015,Trif2015} however, not by means of the waiting-time distribution.

Furthermore, a system consisting of JJ and resonator makes it generally possible to observe both photonic and current statistics (\emph{Josephson photonics}). Of course, the dynamics of the charge transfer processes can indirectly be investigated by observing the created light. However, it would be natural to also directly consider the waiting-time distribution for subsequent charge transfer processes at the JJ or even a mixed waiting-time distribution relating CP transport statistics to photonic emission events.\\

\begin{acknowledgments}
The authors thank A. Armour, F. Portier, C. Flindt, and D. Dasenbrook for valuable discussions. 
Financial support by the Deutsche Forschungsgemeinschaft (DFG) through AN336/6-1 and the 
SFB/TRR21 and by the IQST is gratefully acknowledged.
\end{acknowledgments}

\bibliography{references}

\end{document}